\documentclass[aps,prl,preprint,showpacs]{revtex4}
\usepackage{amsfonts,latexsym}
\begin{document}
\title{Modulo Three Problem With A Cellular Automaton Solution}
\author{Hao Xu, K. M. Lee and H. F. Chau\footnote{Corresponding author,
 Email address: hfchau@hkusua.hku.hk}}
\affiliation{Department of Physics, University of Hong Kong, Pokfulam Road, Hong
 Kong}
\date{\today}
\begin{abstract}
An important global property of a bit string is the number of ones
in it. It has been found that the parity (odd or even) of this number
can be found by a sequence of deterministic, translational invariant
cellular automata with parallel update in succession for a total
of $O(N^2)$ time. In this paper, we discover a way to check if
this number is divisible by three using the same kind of cellular
automata in $O(N^3)$ time. We also speculate that the method described
here could be generalized to check if it is divisible by four and other
positive integers.
\end{abstract}
\medskip
\maketitle

Cellular automaton is a simple-minded discrete dynamical system that
exhibits complex behavior collectively. It has been used in the study
of pattern formation \cite{pf1,pf2} and
many other physical systems \cite{fluid,elchem}, for example.
In order to fully explore the capability of cellular automata, it is
instructive to investigate if and how they can perform some ``simple''
tasks. It is known that cellular automata can do 
a number of basic computations such as finding the
majority as well as the parity of an arbitrary bit
string \cite{majority,parity}, as follows. Let $\sigma$ be an arbitrary bit
string of length $N$ with periodic boundary conditions, $S_\sigma$
be the number of $1$s in $\sigma$. The majority problem is to ask
whether $S_\sigma$ is greater than $N/2$, while the so-called parity
problem is to ask if $S_\sigma$ is odd or even. It has been discovered
that these two problems can be solved by applying a number of cellular
automaton (CA) rules in successions. These two problems (and solutions)
illustrate one important point that global properties of a bit
string can be found by cellular automata, which only know the local
states of the bit string and do not have any memory.

In this paper, we
would like to find out if $S_\sigma \equiv 0 \pmod{3}$. More precisely, we
would like to evolve the given input bit string with periodic
boundary conditions using a sequence
of translational invariant and deterministic CA rules with parallel update to
$0^N$ if $S_\sigma \equiv 0 \pmod{3}$ and $1^N$ otherwise,
where $0^N$ denotes a string of $N$ consecutive zeros and $1^N$ denotes
a string of $N$ consecutive ones.

The plan of solving this modulo three problem follows the plan
of solving the parity problem. We will first evolve the bit string in
$O(N^3)$ steps to put it in some standard forms. Then, we can
tell apart those standard forms by a sequence of CA rules, and hence
if $S_\sigma \equiv 0$. (In this paper, all equivalences are modulo three.)
However, since there are more cases
to be considered in this problem, we have to use a few more CA rules.

The first CA rule that we need is
\begin{center}
  $R_1$:
    \begin{tabular}{cc}
       1000 & 0001 \\ \hline
       111? & ??11
    \end{tabular}~.
\end{center}
The notation means that if part of the bit string is $1000$,
then in the next time step, the first bit of this part remains $1$, the
second and the third bits become $1$. The outcome of the fourth bit
also depends on the bit pattern on its right. Similarly, if some part of the
bit string is $0001$, then the fate of the first and the second bit
depends on the bit pattern on their left, and the third and fourth
bits are $1$. If the local configuration of the bit string is neither
$1000$ nor $0001$, the bits remain the same. For example, under this
rule, $1000001$ becomes $1110011$, $10001$ becomes $11111$ and
$10010101$ remains unchanged. It is straight-forward to check that this is
a $r=2$ CA rule. That is, the step of each bit in the next time step depends
on the state itself as well as its nearest and second-nearest neighbors.

The function of this rule is as follows. If the number of consecutive
zeros in the string is more than three, then the rule will replace the
three ending zeros, two on the left and one on the right, by ones.
Since we are using periodic boundary
conditions, the number of ones modulo three does not change. If we apply
the rule $\lfloor N/3\rfloor$ times, then there will be no three or more
consecutive zeros in a row.

This is a good time to introduce a concept we called the partition
number. Apart from the configurations $0^N$ and $1^N$, groups of
zeros are separated by groups of ones, and the number of groups of zeros
is equal to the number of groups of ones. We define the partition number
of a bit string to be the number of groups of zeros, if the bit string
is not $0^N$ or $1^N$. The partition numbers of $0^N$ and $1^N$ are
defined to be zero. For example, the partition number of
$100111001011$ is equal to $3$ (because of the periodic
boundary condition). The partition number will not increase after
we apply the CA rule $R_1$.

For an arbitrary bit string $\sigma$, $R_1^{\lfloor N/3\rfloor}\sigma$
could be $1^N$ or $0^N$. If this is not the case, then it
does not have three or more consecutive zeros in a row, but there could
be single or two zeros separated by ones. The next thing that we need to
do is to merge the zeros.
\begin{center}
  $R_2$:
    \begin{tabular}{c}
       101 \\ \hline
       01?
    \end{tabular}~.
\end{center}
Again, it is straight-forward to check that $R_2$ is a $r=1$ CA rule.
This rule moves the single zeros to the left. Hence,
the number of zeros is conserved under this rule. Applying this rule
$N$ times,
\begin{equation}
  R_2^N R_1^{\lfloor N/3\rfloor}\sigma~,
\end{equation}
there will be either no single zeros or all the zeros are single. Similar
to $R_1$, the partition number will not increase after we apply $R_2$.

If we repeat the whole process $N$ times, it is straight-forward to see that
the resultant bit string,
\begin{equation}
  (R_2^N R_1^{\lfloor N/3\rfloor})^N\sigma~,
\end{equation}
will have the following properties:
\begin{enumerate}
\item It is equal to $1^N$ or $0^N$. If not, then all the zeros are either
      single or in pairs. There is no three
      or more consecutive zeros, and there will not be both single and
      paired zeros at the same time.
\item The number of ones will be equivalent to $S_\sigma$ modulo three.
\item The partition number of the resultant bit string will be less than
      or equal to the partition number of $\sigma$.
\end{enumerate}

The roles of the zeros and ones could be reversed and we define the CA rules
\begin{center}
  $R_3$:
    \begin{tabular}{cc}
       0111 & 1110 \\ \hline
       000? & ??00
    \end{tabular}~,
  $R_4$:
    \begin{tabular}{c}
       010 \\ \hline
       10?
    \end{tabular}~.
\end{center}

\medskip\noindent\emph{Lemma~1}. Let
\begin{equation}
  \sigma_1\equiv \left[ \left(R_4^N R_3^{\lfloor N/3\rfloor} \right)^N
    \left(R_2^N R_1^{\lfloor N/3\rfloor} \right)^N\right]^N\sigma~,
\label{sigmaone}
\end{equation}
then $\sigma_1$ will be one of the following six cases:\\
A. $1^N$;\\
B. $0^N$;\\
C. $(10^{3n+1})^m$;\\
D. $(110^{3n+2})^m$;\\
E. $(10^{3n+2})^m$;\\
F. $(110^{3n+1})^m$.

\noindent\emph{Proof}. Since $R_3$ and $R_4$ are obtained by interchanging
the roles of $0$ and $1$ in $R_1$ and $R_2$ respectively, our earlier
discussions imply that all ones in $\sigma_2 \equiv (R_4^N R_3^{\left\lfloor
N/3\right\rfloor} )^N ( R_2^N R_1^{\left\lfloor N/3\right\rfloor} )^N \sigma$
are either single or in pairs provided that $\sigma_2 \neq 1^N$. Besides,
$\sigma_2$ does not contain both single and paired ones in this case. That is
to say, $\sigma_2 = 1^i 0^{3n_1+j_1} 1^i 0^{3n_2+j_2} 1^i 0^{3n_3+j_3} \cdots$
up to a linear translation where $i= 1$ or 2, $j_k=0,1$ or 2 and
$n_k \in {\mathbb N}$ for $k$ to run from $1$ to the partition number of
$\sigma_2$.

It is straight-forward to check that the partition number of $\sigma_3
= (R_2^N R_1^{\left\lfloor N/3 \right\rfloor} )^N \sigma_2$ equals that of
$\sigma_2$ if and only if $j_k = j_{k'} \neq 0$ for all $k,k'$. And in this
case, $\sigma_3 = 0^{j_1} 1^{n_1 + 2n_2 + i} 0^{j_1} 1^{n_2 + 2n_3 + i} 0^{j_1}
\cdots$ up to a linear translation. By the same token, the partition number of
$\sigma_4 = (R_4^N R_3^{\left\lfloor N/3 \right\rfloor}
)^N \sigma_3$ equals that of
$\sigma_3$ if and only if $n_k + 2 n_{k+1} + i \neq 0 \pmod 3$ and $n_k +
n_{k+1} - 2 n_{k+2} = 0 \pmod 3$ for all $k$. These conditions imply that
$n_k - n_{k+1} = \alpha \pmod 3$ is a constant independent of $k$ and
$\alpha + i \neq 0 \pmod 3$.
In other words, the possible values of $(\alpha,i)$ are $(0,1)$, $(0,2)$,
$(1,1)$ and $(2,2)$.
Furthermore, it is easy to see that if the partition numbers of $\sigma_3$ and
$\sigma_4$ agree, then up to a linear translation,
\begin{equation}
\sigma_4 = \left\{ \begin{array}{ll}
 1^i 0^{j_1 + (n_1 + 4n_2 + 4n_3)/3} 1^i 0^{j_1 + (n_2 + 4n_3 + 4n_4)/3}
 1^i 0^{j_1 + (n_3 + 4n_4 + 4n_5)/3} \cdots &
 \mbox{if~} \alpha = 0, \\
 1^2 0^{j_1 + (n_1 + 4n_2 + 4n_3)/3 -1} 1^2 0^{j_1 + (n_2 + 4n_3 + 4n_4)/3 -1}
 1^2 0^{j_1 + (n_3 + 4n_4 + 4n_5)/3 -1} \cdots &
 \mbox{if~} \alpha = i = 1, \\
 1 0^{j_1 + (n_1 + 4n_2 + 4n_3)/3 +1} 1 0^{j_1 + (n_2 + 4n_3 + 4n_4)/3 +1}
 1 0^{j_1 + (n_3 + 4n_4 + 4n_5)/3 +1} \cdots &
 \mbox{if~} \alpha = i = 2. \\
\end{array}\right. \label{E:sigma4}
\end{equation}

{}From the above discussions, we know that if the partition number of
$\sigma_4$ equals that of $\sigma_2$ and if not all $n_k$ are equal, then
the action of $( R_4^N
R_3^{\left\lfloor N/3 \right\rfloor})^N (R_2^N R_1^{\left\lfloor N/3
\right\rfloor})^N$ increases the length
of at least one substring of consecutive
zeros (namely, the one with the least value of $n_k$)
while at the same time decreases the length of at least one
substring of consecutive zeros (namely, the one with the maximum value of
$n_k$).

At this point, we introduce the notation of minimum gap
$\ell_{\rm min} (\sigma) = \min \{ n : 10^n1
\mbox{~is~a~substring~of~} \sigma \}$. In other words,
the minimum gap of a bit string is the minimum distance
in between two ones in the string $\sigma$. Clearly, this definition is
well-defined provided that $\sigma \neq 0^N$ or $1^N$. The notation of maximum
gap $\ell_{\rm max}$ is defined in a similar way.

The above discussions imply that if the partition numbers of $\sigma_4$
and $\sigma_2$ agrees, then
\begin{equation}
\ell_{\rm min} ( \sigma_4 ) \geq \ell_{\rm min}
(\sigma_2) \label{E:ell_inc}
\end{equation}
provided that $\sigma_4 \neq 0^N$ or $1^N$.
Furthermore, it is not difficult to see that
Eq.~(\ref{E:ell_inc}) also holds when the partition numbers of $\sigma_4$ and
$\sigma_2$ disagree.
In addition, since at least one substring $10^{\ell_{\rm min} (\sigma_2)}1$ of
$\sigma_2$ becomes $10^{\ell'}1$ with $\ell'>\ell_{\rm min} (\sigma_2)$
under the action of $(R_4^N R_3^{\left\lfloor N/3
\right\rfloor})^N (R_2^N R_1^{\left\lfloor N/3 \right\rfloor})^N$
provided that not all $n_k$ are equal, we conclude
that $\ell_{\rm min}$ of a bit string increases
by at least 1 under the action of $[(R_4^N R_3^{\left\lfloor N/3
\right\rfloor})^N (R_2^N R_1^{\left\lfloor N/3 \right\rfloor})^N ]^{m-1}$
where $m$ is the partition number until the string becomes the forms A to F
in the lemma.

To reach any one of these six forms from an arbitrary bit string $\sigma$,
at most $k$ applications of $(R_4^N R_3^{\left\lfloor N/3 \right\rfloor})^N
(R_2^N R_1^{\left\lfloor N/3 \right\rfloor})^N$ is required where
$k \leq 1 + (m-1) \ell_{\rm max} (\sigma_2) \leq N$. Hence, the lemma
is proved. \hfill$\Box$
\medskip

\begin{table}
\begin{tabular}{|c|c|c|c|c|}
\hline
                   & $N\equiv 1$ & $N\equiv 2$ & $N=3q$ & $N=3^pq$ \\ \hline
$S_\sigma\equiv 0$ & B           & B           & A,B,C,D& A,B,C,D,E,F \\ \hline
$S_\sigma\equiv 1$ & A           & C,D         & E,F    & E,F \\ \hline
$S_\sigma\equiv 2$ & C,D         & A           & E,F    & E,F \\ \hline
\end{tabular}
\caption{This table summarizes the results deduced from the lemma. The
         equivalent relations are those of modulo three. Hence, for
         example, $N\equiv 1$ means $N$ could be 4, 7, 10, etc. $q$ is
         not divisible by three and $p>1$.}
\label{Thetable}
\end{table}

By the lemma, we can easily deduce some preliminary results, which are
summarized in Table~\ref{Thetable}. If $N \equiv 1 \pmod{3}$, then the
case~E and F cannot occur, because the number of bits in case~E and F
are $(3n+3)m$, which is equivalent to zero modulo three. Similarly,
we find that the number of bits in case~A is $N$, which is
equivalent to one modulo three, hence, $S_\sigma\equiv 1$. For case~B,
$S_\sigma\equiv 0$; while for case~C or D, $S_\sigma\equiv 2$.

The other cases in the table can be analysed in a similar way. For example,
if $N=3q$ where $q$ is not divisible by three, and if $S_\sigma\equiv 0$,
then it cannot be case~E. Otherwise $N=(3n+3)m=3(n+1)m$ implies that $m$
is not divisible by three, but $S_\sigma\equiv m$. We have a contradiction.

To reach our goal of determining whether $S_\sigma\equiv 0$ is easy
for $N\equiv 1$ or $2$. Consider the $r=1$ CA rule
\begin{center}
  $R_5$:
    \begin{tabular}{c}
       10 \\ \hline
       11
    \end{tabular}~.
\end{center}
By applying this rule $N$ times, we could turn all the zeros to ones, if
there is at least an one initially. In summary, we have the following.

\medskip\noindent\emph{Theorem 1}. If $N\equiv 1,2 \pmod{3}$,
then for any bit string $\sigma$ of $N$ bits,
\begin{equation}
  R_5^N\left((R_4^N R_3^{\lfloor N/3\rfloor})^N
    (R_2^N R_1^{\lfloor N/3\rfloor})^N\right)^N \sigma
  = \left\{\begin{array}{ll}
             0^N & \mbox{if $S_\sigma\equiv 0$,} \\
             1^N & \mbox{if $S_\sigma\equiv 1,2$.}
           \end{array}\right.
\end{equation}
\medskip

We can, in fact, completely determine $S_\sigma$ modulo three, because
it could be easily constructed a sequence of CA rules which keep case~A
and B unchanged and evolve case~C and D to alternating ones and zeros.
Then, we could obtain the desired result by reading any two or three
consecutive bits. Details will be reported elsewhere \cite{thesis}.

For the case $N=3q$ or $3^pq$, we have to do more work. Let us
first consider the case $N=3q$. Define
\begin{center}
  $R_6$:
    \begin{tabular}{cc}
       10 & 1100\\ \hline
       11 & 1111
    \end{tabular}~.
\end{center}
Notice that this is indeed a CA rule, in the sense that the two
instructions stated in the rule are consistent with each other.
Apply it once, case~A and B remain unchanged, and
\begin{eqnarray}
  R_6 (10^{3n+1})^m  &=& (110^{3n})^m \\
  R_6 (110^{3n+2})^m &=& (11110^{3n})^m \\
  R_6 (10^{3n+2})^m  &=& (110^{3n+1})^m \\
  R_6 (110^{3n+1})^m &=& (11110^{3n-1})^m~.
\end{eqnarray}
Applying $R_1^{\lfloor N/3\rfloor}$ to the result, we can turn case~C and
D to $1^N$, while for case~E and F, we have a bit sting with both ones and
zeros. Therefore, we have
\begin{eqnarray}
  && R_1^{\lfloor N/3\rfloor}R_6\left((R_4^N R_3^{\lfloor N/3\rfloor})^N
     (R_2^N R_1^{\lfloor N/3\rfloor})^N\right)^N \sigma
     \nonumber \\
 &=& \left\{\begin{array}{ll}
             0^N \mbox{ or } 1^N & \mbox{if $S_\sigma\equiv 0$} \\
             \mbox{a bit string with both ones and zeros}
             & \mbox{if $S_\sigma\equiv 1,2$}
           \end{array}\right.
\end{eqnarray}
for any bit string $\sigma$ of length $3q$. To put the resulting bit
string to the final form, we need
\begin{center}
  $R_7$:
    \begin{tabular}{cc}
       11 & 01\\ \hline
       ?0 & ?1
    \end{tabular}~.
\end{center}
Applying this rule once, $0^N$ remains unchanged, $1^N$ becomes $0^N$ and
a bit string with both ones and zeros will become another also with
both ones and zeros. Hence, we have the following theorem.

\medskip\noindent\emph{Theorem 2}. If $N=3q$, where $q$ is not divisible
by three, then for any bit string $\sigma$ of $N$ bits,
\begin{eqnarray}
  &&R_5^NR_7R_1^{\lfloor N/3\rfloor}R_6\left((R_4^N R_3^{\lfloor N/3\rfloor})^N
     (R_2^N R_1^{\lfloor N/3\rfloor})^N\right)^N \sigma
     \nonumber \\
 &=& \left\{\begin{array}{ll}
             0^N & \mbox{if $S_\sigma\equiv 0$} \\
             1^N & \mbox{if $S_\sigma\equiv 1,2$}
           \end{array}\right.~.
\end{eqnarray}
\noindent\emph{Proof}. We just have to spell out the final steps,
\begin{eqnarray}
  &&R_5^NR_7R_1^{\lfloor N/3\rfloor}R_6\left((R_4^N R_3^{\lfloor N/3\rfloor})^N
     (R_2^N R_1^{\lfloor N/3\rfloor})^N\right)^N \sigma
     \nonumber \\
 &=& R_5^NR_7
     \left\{\begin{array}{ll}
             0^N \mbox{ or } 1^N & \mbox{if $S_\sigma\equiv 0$} \\
             \mbox{a bit string with both ones and zeros}
             & \mbox{if $S_\sigma\equiv 1,2$}
           \end{array}\right. \nonumber \\
 &=& R_5^N
     \left\{\begin{array}{ll}
             0^N & \mbox{if $S_\sigma\equiv 0$} \\
             \mbox{a bit string with both ones and zeros}
             & \mbox{if $S_\sigma\equiv 1,2$}
           \end{array}\right. \nonumber \\
 &=& \left\{\begin{array}{ll}
             0^N & \mbox{if $S_\sigma\equiv 0$} \\
             1^N & \mbox{if $S_\sigma\equiv 1,2$}
           \end{array}\right.~.
\end{eqnarray}
\hfill$\Box$

\medskip
Finally, we consider the most difficult case, where $N=3^pq$, $p>1$ and
$q$ is not divisible by three. Because the proof of the theorem is not
very illuminating, we will be brief here. Define $\sigma_1$ by
\begin{equation}
  \sigma_1\equiv \left[ \left(R_4^N R_3^{\lfloor N/3\rfloor} \right)^N
    \left(R_2^N R_1^{\lfloor N/3\rfloor} \right)^N\right]^N \sigma~,
\label{sigmatwo}
\end{equation}
We will evolve $\sigma_1$ to $0^N$ or $1^N$
if $S_\sigma\equiv 0$ and to a bit string with both ones and zeros
if $S_\sigma\equiv 1,2$.

Define $\sigma_2$ be
$R_3^{\lfloor N/3\rfloor}R_1^{\lfloor N/3\rfloor}R_6\,\sigma_1$. It
could be easily checked that $1^N$ and $0^N$ remain unchanged. The case~C
and D will be turned to $1^N$, while case~E and F will switch to each
other. We have taken care of cases~C and D. They will not appear again
in the following discussion.

Define
\begin{center}
  $R_8$:
    \begin{tabular}{cc}
       0100 & 110 \\ \hline
       0111 & 111
    \end{tabular}~,
  $R_9$:
    \begin{tabular}{c}
       10 \\ \hline
       01
    \end{tabular}~.
\end{center}
$R_9$ is the so-called traffic rule, it is just the Wolfram
elementary CA rule 184 \cite{trafficrule}.
Note that $R_8 (10^{3n+2})^m=R_8 (110^{3n+1})^m
=(1110^{3n})^m$. Hence, we have merged the case~E and F. If $n=0$, then
$m$ must be divisible by three, and after we apply $R_8$, we have $1^N$.
This is exactly what we want. Now, we assume $n>0$ and apply $R_9$ twice,
we have $R_9^2(1110^{3n})^m = (101010^{3n-2})^m$. Similar to the
argument we have given in the proof of the lemma, if we apply
$(R_{11}^{\lfloor N/2\rfloor}
R_{10}^{\lfloor N/2\rfloor})^{\lfloor N/2\rfloor}$ to $(101010^{3n-2})^m$,
we get $(10^n)^{3m}$, where
\begin{center}
  $R_{10}$:
    \begin{tabular}{cc}
       1000 & 001 \\ \hline
       110? & 011
    \end{tabular}~,
  $R_{11}$:
    \begin{tabular}{cc}
       011 & 1110 \\ \hline
       001 & ?100
    \end{tabular}~.
\end{center}
Finally, we check that
\begin{eqnarray}
  &&R_3^{\lfloor N/3\rfloor}R_1^{\lfloor N/3\rfloor}R_6
    R_3^{\lfloor N/3\rfloor}R_1^{\lfloor N/3\rfloor}\, (10^n)^{3m}\nonumber \\
 &=& \left\{\begin{array}{ll}
             (110^{n-1})^{3m} & \mbox{if $n+1$ is divisible by three} \\
             1^N & \mbox{otherwise}
           \end{array}\right.~.
\end{eqnarray}
In the first case, we are back to case~F in the lemma, but now, with
$m$ replaced by $3m$. While in the second case, since $n+1$ is not
divisible by three and $N=3m(n+1)$, we have what we expected:
$S_\sigma\equiv 0$.

We have to repeat $p-2$ more times to test whether the original $m$
is divisible by three, hence we have the following theorem.

\medskip\noindent\emph{Theorem 3}. If $N=3^pq$, where $p>1$
and $q$ is not divisible by three,
then for any bit string $\sigma$ of $N$ bits, define $\sigma_1$ by
Eq.~(\ref{sigmatwo}), we have
\begin{eqnarray}
  &&\left(R_3^{\lfloor N/3\rfloor}R_1^{\lfloor N/3\rfloor}R_6
          R_3^{\lfloor N/3\rfloor}R_1^{\lfloor N/3\rfloor}
          (R_{11}^{\lfloor N/2\rfloor}
          R_{10}^{\lfloor N/2\rfloor})^{\lfloor N/2\rfloor}
          R_9^2R_8\right)^{p-1}
    R_3^{\lfloor N/3\rfloor}R_1^{\lfloor N/3\rfloor}R_6\,\sigma_1
    \nonumber \\
 &=& \left\{\begin{array}{ll}
             0^N \mbox{ or } 1^N & \mbox{if $S_\sigma\equiv 0$} \\
             \mbox{a bit string with both ones and zeros}
             & \mbox{if $S_\sigma\equiv 1,2$}
           \end{array}\right.~.
\end{eqnarray}

\medskip\noindent\emph{Corollary~1}. If we apply $R_5^NR_7$ to the
resulting bit string
of Theorem~3, we will have $0^N$ if $S_\sigma\equiv 0$ and $1^N$ if
$S_\sigma\equiv 1,2$.
\medskip

Together with this corollary, we have provided a solution to the
problem we stated at the beginning of this paper, namely, we could
evolve any given bit string with periodic boundary conditions using
a sequence of translational invariant and deterministic CA rules with
parallel update to
$0^N$ if $S_\sigma \equiv 0 \pmod{3}$ and $1^N$ otherwise, and we could
do this in $O(N^3)$ time steps.

We believe that there is no difficulty, in principle, to generalize
the above result to calculate $S_\sigma$ modulo four, five, or other
bigger integers, although we expect the actual implementation will
be tedious and complicated. If we calculate $S_\sigma$ modulo $k$,
we have found that the worst case run time for our algorithms scales
as $O(N^k)$ for $k=2,3$. We do not know yet if this worst case run
time estimation holds for larger $k$, or it will saturate at $O(N^3)$.
We also do not know if there is other more efficient method.
Either way, we should not underestimate
the full power of cellular automata. We have shown just one of their
arithmetic capability and other possibilities are under investigations.

\begin{acknowledgments}
This work is supported in part by the Hong Kong SAR Government RGC grant
HKU~7098/00P. H.F.C. is also supported in part by the Outstanding Young
Research Award of the University of Hong Kong.
\end{acknowledgments}

\end{document}